# MolecularWebXR: Multiuser discussions about chemistry and biology in immersive and inclusive VR

**Authors:** Fabio J. Cortés Rodríguez,[1] Gianfranco Frattini,[2] Fernando Teixeira Pinto Meireles,[1] Danaé A. Terrien,[1] Sergio Cruz-León,[3] Matteo Dal Peraro,[1] Eva Schier,[1] Diego M. Moreno[2] and Luciano A. Abriata[1*]

[1] School of Life Sciences, École Polytechnique Fédérale de Lausanne, CH-1015 Switzerland.
[2] Instituto de Química Rosario (IQUIR, CONICET-UNR) and Facultad de Ciencias Bioquímicas y Farmacéuticas, Universidad Nacional de Rosario, Rosario, Santa Fe, Argentina.
[3] Department of Theoretical Biophysics, Max Planck Institute of Biophysics, Frankfurt am Main, Germany.

* Corresponding autor: luciano.abriata@epfl.ch

**Abstract:** MolecularWebXR is our new website for education, science communication and scientific peer discussion in chemistry and biology built on WebXR. It democratizes multi-user, inclusive virtual reality (VR) experiences that are deeply immersive for users wearing high-end headsets, yet allow participation by users with consumer devices such as smartphones, possibly inserted into cardboard goggles for immersivity, or even computers or tablets. With no installs as it is all web-served, MolecularWebXR enables multiple users to simultaneously explore, communicate and discuss chemistry and biology concepts in immersive 3D environments, manipulating objects with their bare hands, either present in the same real space or scattered throughout the globe thanks to built-in audio features. A series of preset rooms cover educational material on chemistry and structural biology, and an empty room can be populated with material prepared *ad hoc* using moleculARweb's VMD-based PDB2AR tool. We verified ease of use and versatility by users aged 12-80 in entirely virtual sessions or mixed real-virtual sessions at science outreach events, student instruction, scientific collaborations, and conference lectures. MolecularWebXR is available for free use without registration at https://molecularwebxr.org, and a blog post version of this preprint with embedded videos is available at https://go.epfl.ch/molecularwebxr-blog-post.

## Introduction

In both education and everyday work, the chemical and biological sciences rely heavily on the human ability to visually grasp and communicate the details of objects that are inherently three-dimensional, yet most technical means of presenting and manipulating 3D information are intrinsically two-dimensional.[1] Today, most molecular visualization and manipulation takes place in the form of two-dimensional interfaces and representations such as pictures, diagrams and flat-screen computer graphics. Even the most advanced molecular graphics programs display molecules in flat 2D screens at their core, and they enable user interaction with the molecular system only through mouse moves and key strokes which are inherently one-handed and very limited in terms of natural interactivity, let alone in allowing concurrent action of multiple users on the same systems.

To alleviate these drawbacks, over various "waves of hype" on virtual reality (VR) most classical programs for molecular modeling and molecular graphics have introduced VR extensions that provide immersive visualization and spatial control on manual operations.[2–4] In addition, various programs have been developed specifically for manipulating molecules in VR, many with limited functionality compared to the VR versions of more complete molecular graphics software but at the expense of being much simpler to deploy and utilize. Besides, tests have been conducted to explore the benefits and drawbacks of the technology in real-world usage, including tests on how multiple users can work concurrently in shared VR sessions.[5] However, the adoption of VR software for molecular graphics has been very limited, and one could argue that the lack of success is most likely due to a technology that was immature until this decade.

Recently, a new wave of VR, and also augmented reality (AR), technologies has opened up new frontiers for immersive and interactive learning experiences and is now among the fastest evolving consumer product technologies.[6] This renewed interest, along with the substantial investment fueled by the idea of building the pieces of a "Metaverse", has fostered the development of new AR and VR hardware. Newer devices are smaller and more wearable, have sensors that track the environment and the user's pose with high accuracy, and are fully built into the headset eliminating the need to connect to external computers. Recent AR and VR headsets include built-in Wi-Fi connectivity and web browsing capabilities and are more affordable. For example, immersive devices like Meta's Oculus Quest 2, used here, are available for less than USD 400 (November 2023). With the advent of this new AR and VR technology, new programming standards have evolved coordinated by major vendors, and new software tools have emerged in the last years that allow users to teach, self-learn and work in very immersive ways.[7] In the specific context of chemistry and biology,[8] stand-alone programs for AR/VR and modules that extend the capabilities of regular computer graphics programs into AR-VR have emerged.[9–14] Importantly, the most advanced of these AR/VR programs are starting to support at least two-user sessions,[5] paving the way for tools that enable virtual human interactions and collaborations as required for teaching, collaborative work and discussions, etc.

Despite the recent emergence of several new programs for immersive molecular visualization and modeling in VR/AR, their adoption still seems very limited. We argue that the main limitation is not only the cost of the AR and/or VR headsets, which has lowered but is still substantial, but also involves the complex nature of the setups involved and the limited cross-device compatibility of the existing programs. This is why we have been advocating over the last 5 years for purely web-based solutions, especially capitalizing on the WebXR standard and API.[15] WebXR-based solutions bring two advantages of unparalleled relevance to adoption and democratization of access to immersive content: (i) supreme portability across devices and operating systems, from laptops and smartphones to high-end AR/VR headsets; and (ii) zero barriers to deployment and availability as all content runs inside web browsers hence requires no installs or updates and is instantly available upon internet connection.[15–17]

With this philosophy in mind, 3 years ago we released moleculARweb, a free platform that offers several activities and playgrounds for chemistry and structural biology education in commodity AR, that is AR that runs on non-specialized devices such as smartphones, computers, and laptops.[18–20] Next, building on the WebXR standard we extended moleculARweb with PDB2AR, a web tool that allows users to create web-based AR and VR sessions for any kind of consumer device. In PDB2AR the virtual objects are generated with VMD, and as such it supports all its "representations" from simple ball-and-stick models and cartoons to isosurfaces useful to represent electronic orbitals or cryo-electron maps.[21]

Here we introduce MolecularWebXR (Figure 1A), our new platform for chemistry and structural biology education that builds on the power of WebXR to deliver multiuser, cross-device and cross-OS compatible virtual sessions. MolecularWebXR offers rooms with pre-built material about a series of topics useful in chemistry and (structural) biology education, and an empty room that users can populate with custom-made objects from moleculARweb's PDB2AR app to create personalized VR sessions for classes, demonstrations, and scientific discussions. Inside MolecularWebXR sessions, users wearing VR headsets are displayed with their heads and hands reflecting their natural poses and moves; they can grab objects to move them and resize them in space, point with their hands in a natural fashion, and communicate with each other by talking naturally directly through the headset's microphone (Figure 1B). Users without access to hardware specialized for VR can still follow the sessions from their laptops, tablets or smartphones. Moreover, the latter supports immersive modes by using cardboard-made goggles (Figure 1C and 1D). Throughout the article we showcase example applications of MolecularWebXR as either entirely virtual sessions or mixed real-virtual sessions, in (i) science outreach days at our institution, (ii) student instruction, (iii) scientific collaboration, and (iv) at conference lectures.

**System**

MolecularWebXR relies on WebXR, an API and specification for web content and web apps to interface with mixed reality hardware and supported by all major vendors, thus easily enabling wide cross-device compatibility.[15] This API automatically parses the device's input capabilities into standardized events and mechanisms that are fed into the web browser, for which the software is written in JavaScript at the core. A server, running in Node.js, centralizes the creation and management of rooms where VR sessions take place. The exact VR devices used in the experiences presented throughout the figures of this article were the Oculus Quest 2 and Oculus Pro with hand-tracking enabled, having also verified use with handheld controls.

We also verified proper working of MolecularWebXR in Oculus Quest 1 (with handheld controls only), Meta Quest 3 (hand-tracking and controls) and HTC Vive Pro (with controls).

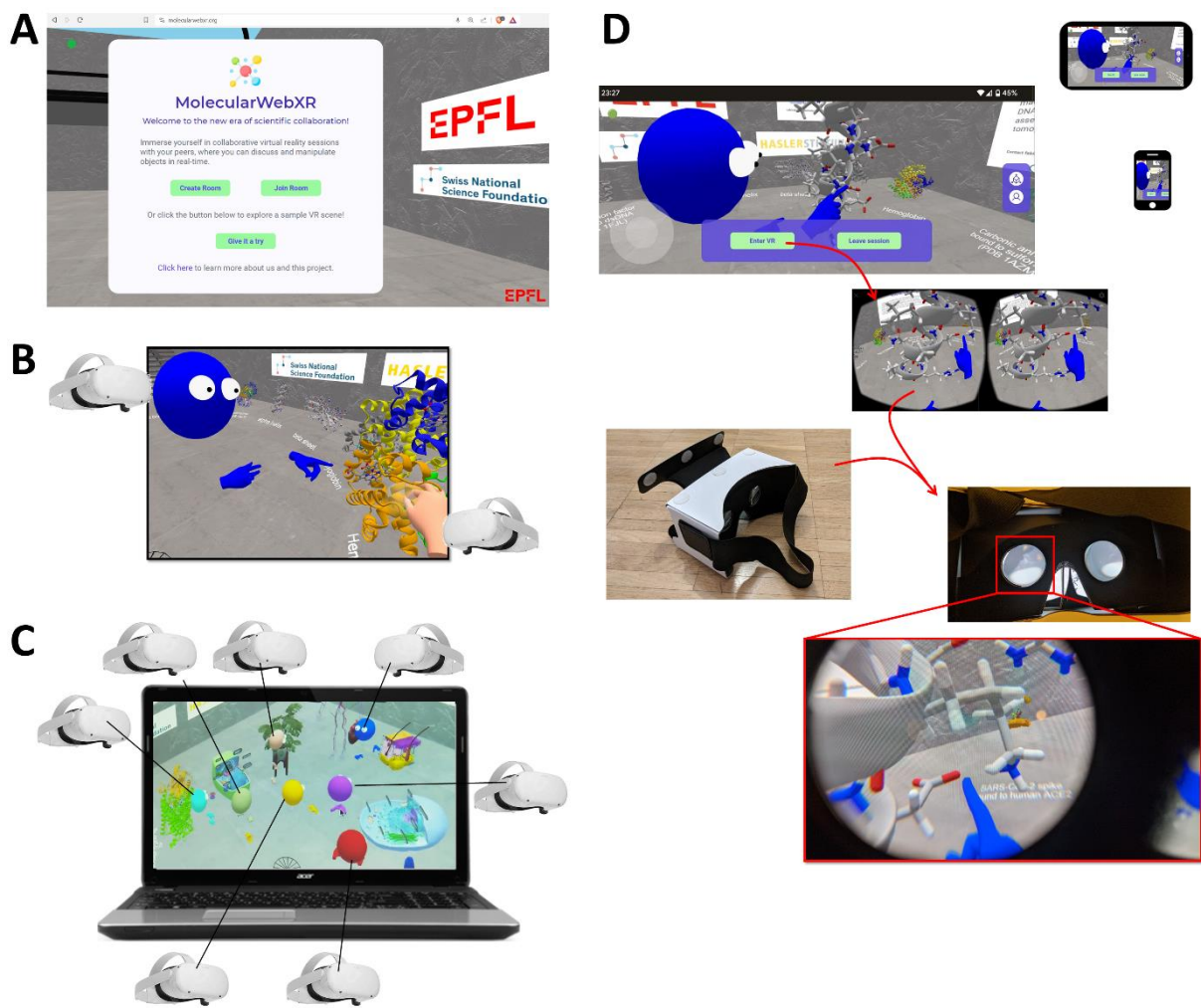

**Figure 1. MolecularWebXR as seen in web browsers on various WebXR-capable devices.** (A) Welcome page as see in a laptop. (B) In VR headsets, the 6 degrees of freedom allow users to explore the VR scene simply by walking around and moving their heads and limbs naturally. In addition, users can translate, rotate and scale objects via natural manual operations with their hands or by using the device's handheld controls. The hands/controls and the heads of participants using VR headsets are broadcast to all other users as simplified avatars. If audio is on, these users can naturally talk to each other and with guests who follow the session from any device can listen to the speakers in VR. (C) A session featuring seven speakers wearing VR headsets as seen from a laptop (looks similar on a tablet). Users can move around the scenes by using the W, A, S and D or the arrow keys if the keyboard (in computers) or a virtual joystick (in smartphones and tablets), and they can direct their looks by using the mouse (computers) or *via* touch gestures (tablets). (D) A session with a user in VR mode accessing through a headset (blue avatar) as seen by a user accessing the session with a smartphone. Outside of VR (top of the panel)s users accessing via smartphones can move around the scenes by using a virtual joystick (Figure S1) and they can gaze *via* touch gestures. In smartphones supporting WebXR (going down through the panel), users can enter WebXR mode and insert the phone into cardboard goggles to explore the scene via 3 degrees of freedom at the entry point, without the ability to move around the session but with the option to move their heads to look around naturally.

By design, access to MolecularWebXR is highly democratized, as the web standard ensures that the software works out of the box in the web browsers of all kinds of devices from high-end VR headsets to smartphones, tablets and computers, leaving no one out (Figure 1B-D). In high-end VR headsets, users can grab objects and pass them around with their hands or controls, zoom them in or out with natural gestures, use their hands to point at objects, and freely move around scenes. Users experiencing VR in headsets are displayed as hand-and-head avatars that all other users can see. In modern smartphones, users can move around the VR scene with a joystick located on the bottom left and choose where to gaze by touching on the screen. They can also access sessions in immersive VR by using cardboard goggles, without the ability to grab objects because smartphones do not (yet) offer built-in hand tracking but with full capability of seeing the full scene and the other users as well as hearing the conversations and talking. Another limitation of smartphones is that since they offer only 3 degrees of freedom, the goggle-assisted VR mode allows for 360-degree visualization but not displacements, which can only be achieved outside of VR mode by using the joystick (Figure 1D, the joystick is the grey circles on the bottom left). Last, in tablets, laptops and other kinds of computers there is naturally no immersive VR of any type; however, users can move around the VR scenes by using the arrow (or W, A, S and D) keys and mouse or touch gestures, as well as seeing, talking to and hearing all other users who are in VR.

**Using MolecularWebXR**

To access the system, users must direct the web browsers of their devices to https://molecularwebxr.org/. On first entry with a given device, users must allow audio functions to support talking over the internet.

Once in the main hall (Figure 1A) a user can create a new room or join an existing one by using a unique code provided by the person who created it, whom we refer to as the *Admin*. When a user creates a new room, it becomes its *Admin* and obtains codes to invite *VR-active* and other users who can act as guests, *i.e.* who can follow the presentation but only passively. *VR-active* users can enter VR if they use a WebXR-capable device and can grab virtual objects with their hands or controls if they are using a VR headset and if the *Admin* has enabled object grabbing. *VR-active* users can also talk to each other and to the *Admin*. All users can move around the VR scene and listen to the *Admin* and *VR-active* users, but those in passive mode cannot grab objects or talk. This distinction between different classes of users allows sessions to be set up in different formats; for example, as discussions where *VR-active* users can present and discuss while other users follow passively; or as courses/lectures/presentations where a single *Admin* or *VR-active* user lectures to various passive users, etc.

Importantly, VR sessions running in MolecularWebXR can accommodate users located in the same physical space or accessing from remote locations (Figure 2A and 2B). In the latter case, the audio features are essential to allow for natural conversation and discussion. If all users are located in the same physical space, the audio features should rather be turned off. For mixed locations, audio features can be kept at the *Admin* level, but ensuring that individuals located in the same physical space turn off all but one of their personal audio inputs.

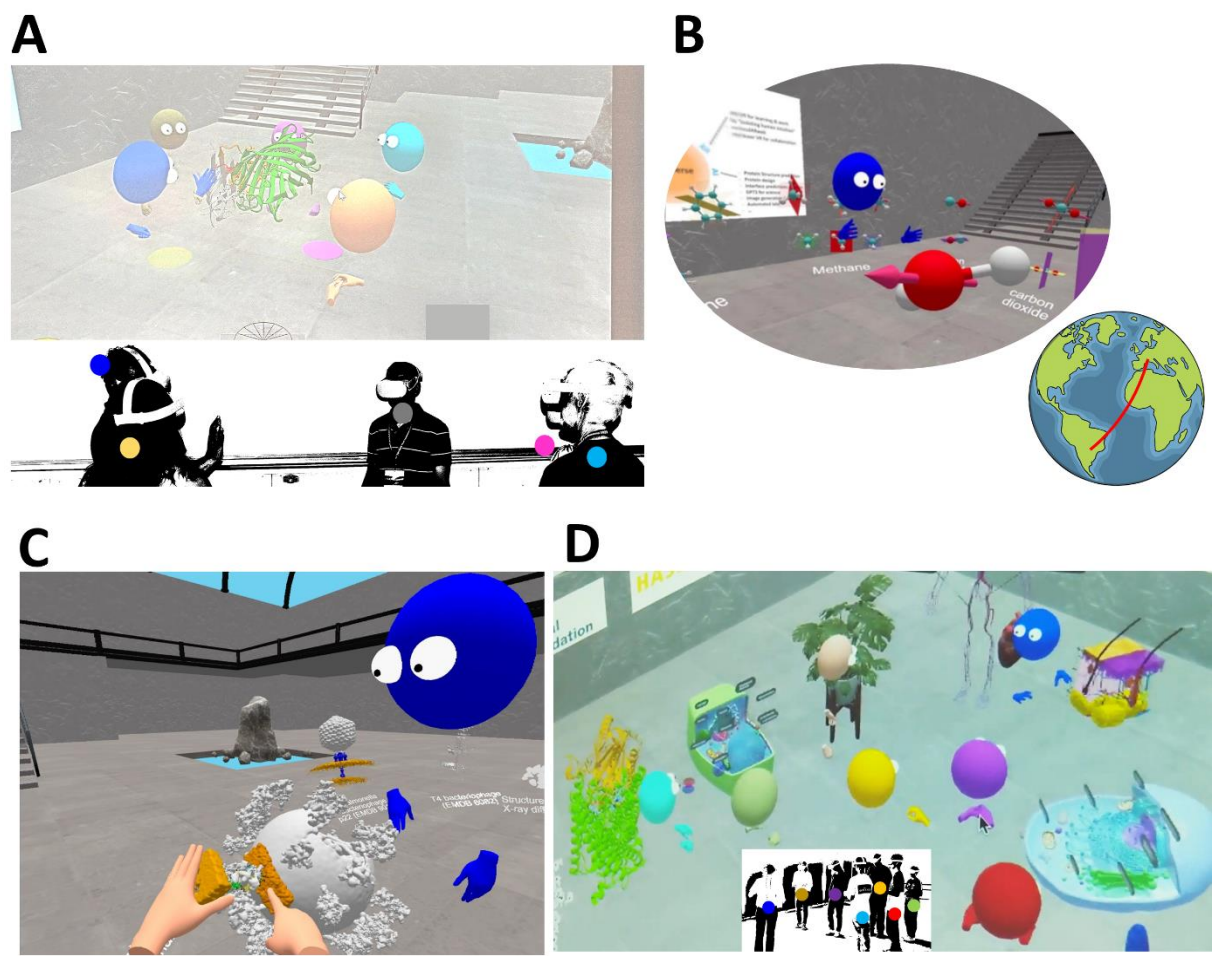

**Figure 2. Real vs. virtual presence in MolecularWebXR, and example content.** (A) Five users attending a VR session where the avatar in blue is describing the structure of a protein complex. Example taken from an application of the website to a subset of talks during a structural biology conference. The safety spaces of the 5 VR headsets were synchronized to optimize match between real and virtual worlds, and audio features were off. Other attendees of the conference could follow the talks by projecting the view of a sixth user accessing through a laptop. In the photograph we have removed colors on the users to mask their identities and have overlaid circles whose colors match the corresponding avatars as seen in the projected view. (B) A teacher in Rosario, Argentina (blue avatar) teaching his students from inside a room populated with VR content about the symmetry elements of molecules, as seen by a visitor accessing in a VR headset from Lausanne, Switzerland, over 11,000 km away. Students in this case followed the teacher's presentation through the view of a third user who accessed the session with a laptop and projected its view on a widescreen. (C) Two users inside the same VR session. View from the orange avatar as she/he is aligning a model of the SARS-CoV2 Spike protein bound to an ACE2 receptor to the electron density of a Spike protein protruding from the viral particle reported from cryo-electron tomography and subtomogram averaging deposited as EMDB 30430[22]. (D) A session deployed over an open science day at EPFL (Lausanne, Switzerland) using MolecularWebXR with seven people in the VR room, on content prepared by combining custom representations of molecules created by PDB2AR or obtained from Sketchfab.com in free or paid forms (this room is not available on the website as it contained purchased VR objects).

It is also important for sessions having users in the same physical space that the safe-space of all VR devices (the "guardians" in the jargon of Oculus/Meta products) be set up in the same way, so that the relative positions of different users match in the real and virtual spaces. With the right setup, users can feel their physical proximity and talk directly to each other in a very natural way, in the same real space but handling the virtual objects and not seeing their real bodies but their avatars.

We note that bandwidth consumption is high when the VR objects are downloaded upon the user entering the room. Once the room is ready, very high-speed Wi-Fi is not needed but it is important to have a stable internet connection to keep updates fluid as the different users move objects and themselves around the virtual room. We note that no video is transmitted between users but just the quaternions that describe object positions, orientations and scales of the VR objects and of the hands and head avatars of all users inside VR.

We have had up to 8 simultaneous *VR-active* users inside VR, an *Admin* running on a laptop and two passive users following the sessions online, without apparent lag (all 8 VR devices and the *Admin* were on the same Wi-Fi network).

**Content and example applications**

When the *Admin* user creates a session, she/he can choose to use a preset room containing content that we have prepared specifically to cover certain topics of chemistry and structural biology expected to benefit largely from deep 3D visualization, or from an empty room where objects can be added at will (Figure 3). Fully customized content can be created via PDB2AR (https://molecularweb.epfl.ch/pages/pdb2ar.html) (Figure 3A). The preset content covers topics on molecular structures(Figure 3B-F), molecular electronic orbitals (Figure 3C), symmetry elements in molecules(Figure 3B), crystal latices and atomic arrangements in simple materials (Figure 3D), introductory structural biology (Figure 3F), and 3D views of cellular compartments obtained by analysis of cryo-electron tomography data (Figure 3G-I).

Immersive 3D visualizations are particularly well suited for analyzing *in situ* cryo-electron tomography data to unravel the spatial interactions that underlie cellular organization. In fact, researchers working at the frontier of automatic structure identification have shown great enthusiasm for the possibilities offered by MolecularWebXR. Indeed, some groups have contributed with cellular 3D reconstructions -and we are open to add more material on request. Figure 3H shows a map of *T. kivui* cells. The model depicts the plasma membrane, S-layer, and the unusual filamentous enzyme that drives $CO_2$ fixation.[23] Figure 3I shows a 3D landscape around the nuclear envelope of *S. pombe* cells, as seen inside VR. The model was generated from a public dataset (EMPIAR: 10988)[24] as described in ref.[25] and displays ribosomes, nuclear pore complexes, fatty acid synthases and the nuclear envelope. Other content in the same room includes models of three different viruses and bacteriophages, with their identifying EMDB codes provided in labels (Figure 3G).

Besides accessing preset content, users can quickly create *ad hoc* content for their presentations starting from either raw PDB files (from the PDB, the AlphaFold-EBI database, or uploaded form the device) or from VMD-generated wavefront objects. In this case, users must follow the procedures described in our previous work,[21] and then copy-paste the link to the GLB file obtained via email. An example application of custom-generated content is shown in Figure 2A for which the users had to prepare virtual representations of the structures they presented as part of a series of scientific talks during a conference.

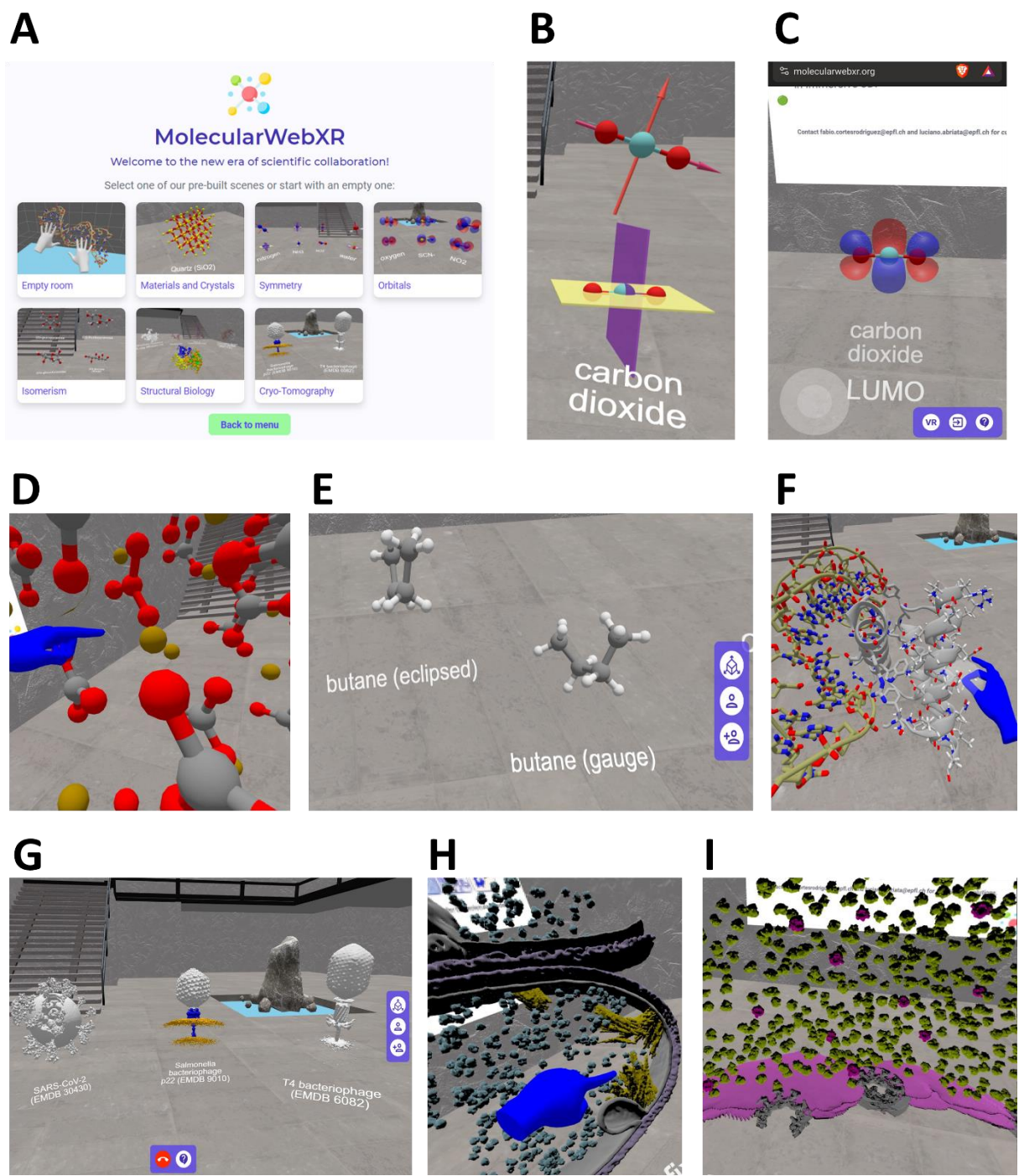

**Figure 3. Examples of 3D content running in VR sessions inside MolecularWebXR.** (A) All the rooms and pre-built content available at MolecularWebXR at launch. Besides the empty room to be populated with objects from PDB2AR, we offer rooms displaying the symmetry elements of molecules from the main point groups, the frontier molecular orbitals most widely studied at university level, examples on isomerism, example structures of materials and crystalline arrangements, example structures of biological macromolecules, and examples of subcellular structures imaged in 3D through cryo-electron tomography. (B) Small cut from a laptop screen around $CO_2$'s symmetry elements, from the room on molecular geometry. (C) Screenshot from a smartphone in portrait orientation showing $O_2$'s LUMO, from the room on molecular orbitals. (D) Zoom inside VR on the local geometry around $Ca^{2+}$ ions in calcite, $CaCO_3$, with the viewer's index finger pointing at a $Ca^{2+}$ ion. (E) Models of butane in eclipsed and gauge conformations, from the room on isomerism. (F) Superimposing by hand an atomically detailed model of an alpha helix onto a helix in a cartoon-only representation of a transcription factor bound to DNA. (G) A SARS-CoV-2 particle, a *Salmonella* bacteriophage punching through the two bacterial membranes (orange) with its needle (blue), and a T4 bacteriophage, all retrieved from the indicated the EMDB entries displayed in the labels, as seen in the room about cryo-tomography for biology. (H) Also from the room on cryo-

tomography for biology, cut of *Thermoanaerobacter kivui* cell showing the carbon-fixing organelles in yellow, the membrane and some of its invaginations in grey, and the S-layer in purple. Model from ref. [23] (I) Another object in the room on cryo-tomography for biology: 3D landscape around the nuclear envelope for a *S. pombe* cell. The model was obtained from ref. [24] as detailed in ref.[25] and shows the nuclear envelope (purple) with some nuclear pore complex subunits (grey), 80S ribosomes (green) and fatty acid synthases (magenta).

Furthermore, starting from the empty room and by leveraging content created in-house with PDB2AR and downloaded or purchased from the 3D art platform Sketchfab.com, we have held sessions designed specifically for certain science communication events and school visits. For example, Figure 2D is a snapshot from a 15 min long VR session where a presenter tells an engaging story that connects physics with chemistry and biology attempting to demonstrate the continuous nature of science, with all participants inside VR.

As described above (Figures 1 and 2), we have successfully applied MolecularWebXR in various formats at scientific presentations in conferences, science outreach events, and for teaching, including combinations where users wore VR headsets or employed other kinds of devices. Importantly, we have had almost 150 people trying the 15 min long VR presentation from Figure 2D and none of the participants had to quit the experience prematurely due to VR sickness or other problems. Furthermore, users who tried the experiences inside VR headsets spanned an age range from 12 to 80 years old, and all of them could seamlessly manipulate objects with their hands, even when they had absolutely no previous experience utilizing VR headsets (>80% of the participants).

**Discussion**

Modern VR and AR headsets allow users to visualize and manipulate 3D models with a level of depth and realism that was unattainable just 5 years ago. Unfortunately, multiple factors hainder widespread access to the technology, leaving out a large proportion of the population as compared to solutions based in regular consumer devices at the expense of being less immersive and more limited in how naturally users can interact with virtual objects. Our previous big release, MoleculARweb, tackled the need for more immersive visualization and more intuitive object control for chemistry and structural biology education through web-based but only partially WebXR-based activities and playgrounds. With a steady base of 2,500 users per month and almost 90,000 users since its launch, MoleculARweb is widely employed at schools. Now, based on similar core concepts and material but building on WebXR and with the aim of achieving more immersive sessions, we hope the new MolecularWebXR will be welcomed for education, science outreach, and scientific discussions. We note that the device we used most for development, testing and application on users, the Meta's Oculus Quest 2 with 128 GB of memory, costs under USD 400 as of November 2023, which is around that of a mid-range smartphone. In October 2023, the Meta Quest 3 was launched for under USD 600, and other companies have their own WebXR-supporting VR devices with prices starting around similar values.

Profiting from the lower costs and building on web standards with cross-device/cross-OS compatibility and ease of deployment as our main values up from the early stages of the project, and by exploiting WebXR's ability to unify all devices, we envision that MolecularWebXR can bridge the digital divide and produce more equal opportunities. It encourages a more inclusive use of VR and democratizes access to the modern possibilities that the technology offers. For example, a well-funded educational institution, company or research lab can afford multiple VR headsets to do concurrent multiuser collaboration or teaching with a large number of students inside immersive VR; while in more modest institutions there could be just one or two users wearing VR headsets to control objects and to present while other users participate from their smartphones, possibly inserted in carboard goggles to feel immersive, or simply on a computer screen or beaming on a widescreen.

We cannot emphasize enough how WebXR makes the experience highly inclusive and content so readily available. We take the chance to encourage other developers creating scientific VR/AR applications to move to this platform. All developers of VR and AR headsets, from major players like Apple, Meta and Microsoft to smaller, highly specialized companies, have included web browsers that support the standard, ensuring that developments for the web should work out of the box in all of them.[15–17] From our side, next in the line is the release of a web app for immersive molecular simulations that will enable educational and research activities like those offered by moleculARweb's virtual modeling kits[19] but in multiuser, immersive VR that runs, like all of MolecularWebXR's content, on VR and non-VR capable devices. A growing universe of tools based on WebXR has the potential to transform the way how chemistry and biology, and science in general, are taught, learned, communicated and discussed, leaving no one behind.

**Acknowledgements**

This work was initiated with funds from Hasler Stiftung (Bern, Switzerland) and then executed with funds from the Swiss National Science Foundation (CRARP2_209794 and 205321_207487) to LAA. We acknowledge Prof. H. Abriel, Dr. P. Teixidor and Dr. V. Rossetti (University of Bern, Switzerland) for their help deploying VR-based scientific presentations during the closing event of NCCR TransCure. We acknowledge Dr. Ricardo Diogo Righetto (University of Basel, Switzerland) for help preparing the 3D view of *T. kivui* cell.